\documentclass[aps,prl,preprint,superscriptaddress]{revtex4}
\usepackage{graphicx}
\usepackage{verbatim}

\begin{document}

L.A. Wray \textit{et.al.}, Nature Physics \textbf{7}, 32 (2011)

News \& Views at http://www.nature.com/nphys/journal/v7/n1/full/nphys1869.html

\title{How robust the topological properties of Bi$_2$Se$_3$ surface are : A topological insulator surface under strong Coulomb, magnetic and disorder perturbations }

\author{L. Andrew Wray}
\affiliation{Department of Physics, Joseph Henry Laboratories, Princeton University, Princeton, NJ 08544, USA}
\author{Su-Yang Xu}
\author{Yuqi Xia}
\affiliation{Department of Physics, Joseph Henry Laboratories, Princeton University, Princeton, NJ 08544, USA}
\author{David Hsieh}
\affiliation{Department of Physics, Massachusetts Institute of Technology, Cambridge, MA 02139, USA}
\author{Alexei V. Fedorov}
\affiliation{Advanced Light Source, Lawrence Berkeley National Laboratory, Berkeley, California 94305, USA}
\author{Hsin Lin}
\author{Arun Bansil}
\affiliation{Department of Physics, Northeastern University, Boston, MA 02115, USA}
\author{Yew San Hor}
\author{Robert J. Cava}
\affiliation{Department of Chemistry, Princeton University, Princeton, NJ 08544, USA}
\author{M. Zahid Hasan}
\affiliation{Department of Physics, Joseph Henry Laboratories, Princeton University, Princeton, NJ 08544, USA}



\date{December, 2010}

\maketitle

\textbf{Three dimensional topological insulators embody a newly discovered state of matter characterized by conducting spin-momentum locked surface states that span the bulk band gap as demonstrated via spin-resolved ARPES measurements  \cite{Intro,TIbasic,MooreandBal,DavidNat1,DavidScience,DavidTunable,hasan,MatthewNatPhys}. This highly unusual surface environment provides a rich ground for the discovery of novel physical phenomena \cite{DavidScience,MacDonaldKerr,FuNew,HorCa, MatthewPreprint, FerroSplitting,FuHexagonal,MatthewTuneBiTe,dhlee,YeHelix,Biswas,
ExcitCapacitor,palee,HorMn,ZhangDyon,KaneDevice,DavidNat1,MatthewNatPhys,
ZhangPred,DavidTunable,YazdaniBack,WrayCuBiSe}. Here we present the first controlled study of the topological insulator surfaces under strong Coulomb, magnetic and disorder perturbations. We have used interaction of iron, with a large Coulomb state and significant magnetic moment as a probe to \textit{systematically test the robustness} of the topological surface states of the model topological insulator Bi$_2$Se$_3$. We observe that strong perturbation leads to the creation of odd multiples of Dirac fermions and that magnetic interactions break time reversal symmetry in the presence of band hybridization. We also present a theoretical model to account for the altered surface of Bi$_2$Se$_3$. Taken collectively, these results are a critical guide in manipulating topological surfaces for probing fundamental physics or developing device applications.}



Bismuth selenide has been discovered by angle resolved photoemission spectroscopy (ARPES) to be a topological insulator with a bulk band gap of about $\sim$300 meV \cite{MatthewNatPhys,DavidTunable, MatthewPreprint}. Spin resolved photoemission studies reveal that surface electrons in Bi$_2$Se$_3$ form a Dirac cone spanning the bulk insulating gap, composed of spin-momentum locked helical states revealing its topological order (Fig. 1a). The Fermi level of grown crystals is usually found to be located in the bulk conduction band due to selenium vacancy defects, however it was subsequently shown that with Ca doping or NO$_2$ surface deposition, the Fermi level can be placed at the Dirac point reaching the topological transport regime \cite{DavidTunable,HorCa} and magnetic interactions can be controlled via Fe or Mn \cite{HorMn, MatthewPreprint}. Placing the Fermi level at the Dirac point in the presence of magnetic impurities in the bulk can lead to a small gap at the Fermi level (Fig. 1b), however the full character of this gap cannot be decisively resolved with the current resolution of ARPES due to lineshape broadening effects. So far, no systematic surface deposition of magnetic impurities to elicit a large systematic magnetic response and bring about controlled changes in the surface band structure has been explored. Probing the effect of magnetic perturbation on the surface is more relevant for potential applications than the previous studies of bulk dopants, because topological insulators need to be in contact with large moment ferromagnets and superconductors for device applications \cite{FerroSplitting,KaneDevice,MacDonaldKerr,Biswas,ExcitCapacitor,palee,ZhangDyon}. The focus of this Letter is the exploration of topological insulator surface electron dynamics in the presence of magnetic, charge and disorder perturbations from deposited iron on the surface.

Bismuth selenide cleaves on the (111) selenium surface plane, providing a homogeneous environment for deposited atoms. Iron deposited on a Se$^{2-}$ surface is expected to form a mild chemical bond, occupying a large ionization state between 2$^+$ and 3$^+$ with roughly 4 $\mu_B$ magnetic moment \cite{IronValence}. As seen in Fig. 1d the surface electronic structure after heavy iron deposition is greatly altered, showing that a significant change in the surface electronic environment has been achieved. Five (odd number of) surface bands intersect the Fermi level rather than just one, and extend below the Fermi level in the form of multiple Dirac cones. Unlike the bulk electronic states of Bi$_2$Se$_3$ \cite{WrayCuBiSe}, these features have no z-axis momentum dependence, confirming that they are largely two dimensional in nature (Fig. 1e).

Figure 2 shows the surface evolution as a function of deposition time in crystals with chemical compositions of (Sample $\#$1) Bi$_{1.9975}$Ca$_{0.0025}$Se$_3$ and (Sample $\#$2) Bi$_2$Se$_3$ varied to tune the carrier density. The rate of iron deposition was similar to 4$\%$ of a monolayer per minute, as detailed in the SI, and deposition data for an additional sample with intermediate carrier concentration (Sample $\#$3) are also presented in the SI. The bulk electronic state of Sample $\#$1 prior to deposition is slightly p-type with significant resistivity ($\rho$ = 23 m$\Omega$cm), and Sample $\#$2 is electron doped from Se vacancy defects in as-grown Bi$_2$Se$_3$. In each case, it is observed that the presence of positively charged Fe surface ions progressively lowers the energy of the surface state and causes the appearance of new surface states with energy/momentum contours similar to the bottom of the bulk conduction band.


The changes brought on by iron deposition can be seen most strikingly in measurements on the bulk p-type Sample $\#$1. After approximately six minutes of deposition, new surface states that do not exist in unperturbed samples become visible at the Fermi level, with new Dirac points labeled D1 and D2 above the Dirac cone that is seen in as-grown Bi$_2$Se$_3$ (D0). It is also after approximately six minutes of deposition that a gap begins to be apparent at the D0 Dirac node, evidenced by a parabolic shape near the Dirac node, separating the upper and lower Dirac cones of the original surface state. This gap can be seen clearly at different incident photon energies in Fig. 2d, confirming that it is a feature of the surface and not the bulk electronic band structure. Electron velocities (band slope) near the D1 and D2 Dirac points increase monotonically as iron is added, showing that iron is increasing the ``Rashba" interaction term (($\vec{k}\times\hat{z})\cdot\vec{\sigma}$, with $\vec{\sigma}$ representing the Pauli matrices) identified in theoretical models \cite{TIbasic}. The number of surface bands intersecting the Fermi level between the $\overline{\Gamma}$- and $\overline{M}$-points progresses from one to three to five, with one band contributed by the original (D0) Dirac cone and two more bands contributed by each of the new (D1,D2) Dirac points. This is consistent with the Mod(2) character of surface electrons on a crystal with bulk topological insulator order, that the topological surface likes to maintain an odd number of Dirac states. After 12 minutes of deposition, the binding energy of the D0 Dirac point was found to have sunk approximately 0.6eV in energy, and the electron binding energies ceased to change under additional deposition. When the chemical potential is positioned above the bulk conduction band minimum, as in Sample $\#$2 (Fig. 2c), the dispersion of new surface states across the full bulk band gap is visible within the photoemission image. A new, strongly split surface band is observed in Sample $\#$2 after five minutes of Fe-deposition with a (D1) Dirac-point at the $\overline{\Gamma}$-point, but no further (e.g. D2) bands appeared after longer deposition.




Theoretical simulation of non-magnetic surface Coulomb perturbation on Bi$_2$Se$_3$ is shown in Fig. 3a, and qualitatively reproduces the progressive appearance of new Dirac points with increasing iron deposition. Through comparison with our numerical result, we can see that the experimentally observed surface states begin to pair off at momentum separation greater than $\sim$0.1$\AA^{-1}$ from the Brillouin zone center, with the upper D0 Dirac cone approaching degeneracy with the lower D1 band, and the upper D1 band connecting to the lower D2 band. The partner-swapping connectivity observed in the simulation and data is a simple way by which new states can be added to the surface band structure without disrupting the surface conditions required by the bulk topological insulator order of Bi$_2$Se$_3$ \cite{TIbasic}. The spin-splitting of topological surface bands is often discussed as a special case of the Rashba effect (e.g. Ref. \cite{TIbasic,FuHexagonal}), in which surface electronic states become spin-split by an energy proportional to their momentum $\vec{k}$. Our data and simulations show that this description is only accurate for Bi$_2$Se$_3$ in a small part of the Brillouin zone surrounding the Brillouin zone center, because at momenta further from the $\overline{\Gamma}$-point the electronic states pair off and are nearly spin degenerate (see Fig. 3b). This can be understood because the origin of the topological insulator state in Bi$_2$Se$_3$ is a symmetry inversion that occurs at the $\Gamma$-point \cite{MatthewNatPhys,ZhangPred}, and the electronic states close to the Brillouin zone boundary are similar to those of topologically trivial materials. The role of magnetic domains in reshaping low energy band structure is expected to be more subtle than Coulomb perturbations, and may be limited to changes near the Dirac points and in self energy lineshape effects (see Fig. 3c-d).


Recent theoretical studies suggest that the physical environment of magnetic impurities on a topological surface is very different from the surface environment provided by a normal semiconductor such as silicon \cite{Biswas,zhangTImagImp,FerroSplitting}. When a non-magnetic crystal is doped with magnetic impurities, long-range magnetic ordering can come about as a result of itinerant electrons exchange-mediating the magnetic interaction. In a normal three dimensional material, electrons that interact with surface-deposited magnetic impurities are free to scatter away from the surface over a 2$\pi$ solid angle, and the two dimensional magnetic interactions therefore typically decay over several Angstroms (e.g. Ref. \cite{adatomExchange}). In topological insulators, surface state electrons are naturally confined to the surface in two dimensional Dirac cones, and it is suggested that interactions between deposited impurities can be mediated over many nanometers (see illustration in Fig. 4c) \cite{Biswas,zhangTImagImp}. For broader theoretical investigations and experiments with nanoscale engineering capabilities, it is interesting to note that exotic magnetic phases (e.g. helical) may occur within certain specific configurations of magnetic impurities such as isolated 1D chains, or from instabilities that arise as the chemical potential is moved far from the surface Dirac point \cite{FuHexagonal,YeHelix}. In the present case of homogeneous 2D deposition with chemical potential near the D0 Dirac point, the momentum-locked spin polarization of topological surface electrons is thought to support ferromagnetic order with an out-of-plane bias ($\vec{B}$ along the $\pm\hat{z}$ direction) at high impurity densities \cite{FerroSplitting,Biswas}, unlike direct dipole-dipole interactions which are much weaker and favor in-plane magnetic orientation.

Our numerical simulations have shown that the new D1 and D2 Dirac point electrons are localized deeper inside of the material than D0, and are expected to interact more weakly with electronic orbitals of the surface-deposited iron. Therefore, we focus on the D0 electrons to understand the effect of iron magnetism on the topological surface. A gap appears at the D0 Dirac point after heavy iron deposition, and the lower D0 Dirac cone acquires a buckled shape with a local energy minimum at zero momentum. The mass induced in the upper D0 Dirac cone after full deposition is approximately 0.1 M$_e$ (electron masses), and the surface band gap is similar to 100 meV. It is difficult to identify the exact time at which the band gap appears, likely because there is no long-range order of iron spins on the surface \cite{MerminWagner,UltrathinMag}, and differently ordered domains will yield differently gapped contributions to the photoemission signal.

Global ferromagnetism breaks time reversal symmetry, making it possible to induce a gap at the Dirac point, which is otherwise disallowed by the crystal symmetry. Figure 4a shows the result of modifying a Generalized Gradient Approximation (GGA) numerical prediction of the D0 surface state by adding perturbative coupling to a surface layer of magnetic impurities with ferromagnetic out-of-plane order (equivalent to the Zeeman effect), yielding a dispersion that closely matches the experimental data. Modifying the surface further with negative-ion deposition could place the Fermi level inside of the surface band gap (Fig. 4a(right)). Details of this calculation and factors that can modify the surface state dispersion are discussed in the SI. If the D0 band gap were due to an isotropic out-of-plane magnetic field, a similar gap would be expected at the D1 and D2 Dirac points, which is clearly not observed. Because magnetic field perturbations from the sub-monolayer iron coating are quite weak, the magnetic phase term in perturbative Hamiltonians can be discounted, and the dominant magnetic effects are expected to be described by a local Heisenberg exchange term representing direct wavefunction overlap (hopping) between surface electrons and spin polarized Fe 3d orbitals \cite{Biswas}. Evidence of Landau levels is neither seen in our data nor in this context expected from theory. The general effect of Zeeman-type magnetic symmetry breaking on topological Dirac surface electrons is shown in Fig. 3c, illustrating that out-of-plane magnetic order will cause a gap, and in-plane magnetic domains will cause bands to shift in momentum space \cite{FerroSplitting}. Based on the appearance of a gap at D0 in our data, we conclude that much of the iron-deposited crystal surface is occupied with domains that have net out-of-plane magnetic moment.

Contrary to the usual trend of deposition experiments, in which photoemission images become increasingly blurry as molecules are haphazardly added to the surface, we identify a regime in which the image becomes qualitatively sharper with the increasing coverage of iron (seen in in Fig. 4d and panels 1-3 of Fig. 2a), corresponding approximately to the increasing clarity of the D0 gap. Momentum-axis width of bands ($\delta$k) was self-consistently measured through Lorentzian fitting at binding energies between the D0 Dirac point and the onset of the D1 band structure. Momentum width is inversely related to the electronic mean free path, and the reduction in $\delta$k as the gap appears is likely indicative of a magnetic disorder-to-order transition, leading to a reduction in scattering. Reducing the prevalence of magnetic domains with in-plane moment could also cause the surface state bands to appear narrower even if electronic mean free paths are unchanged, as can be observed through comparing the domain-averaged band profiles in Fig. 3d. The signature of large ferromagnetic domains with in-plane moment is a twin-peak feature that is not seen in the data. The close correspondence between reduced momentum-distribution-curve (MDC) or $\delta$k width and gapping of the D0 Dirac point therefore strongly indicates the occurrence of an out-of-plane magnetic ordering phase transition after approximately six minutes of iron deposition.

This apparent ordering transition driven by magnetic interactions that are mediated by the topological surface state is an indication of how strong topological insulator order changes the physical environment of the material surface. The observations reported here, including out-of-plane surface magnetism and the appearance of new Dirac surface states, open a window into how topological surface states are formed and interact with various perturbations. As such, they are significant for theoretical understanding of the formation of the topological insulator state in particular materials, and have direct implications for proposed devices utilizing magnetic or Coulomb-charged interfaces with topological insulators, such as magneto-electric junctions \cite{palee,FerroSplitting,KaneDevice,ZhangDyon} and capacitors \cite{ExcitCapacitor} to integrate into multifunctional topological transistors.


\textbf{Methods summary:}
Angle resolved photoemission spectroscopy (ARPES) measurements were performed at the Advanced Light Source beamlines 10 and 12 using 35.5-48 eV photons with better than 15 meV energy resolution and overall angular resolution better than 1$\%$ of the Brillouin zone (BZ). Samples were cleaved and measured at 15$^o$K, in a vacuum maintained below 8$\times$10$^{-11}$ Torr. Momentum along the $\hat{z}$ axis is determined using an inner potential of 9.5 eV, consistent with previous photoemission investigations of undoped Bi$_2$Se$_3$ \cite{MatthewNatPhys, WrayCuBiSe}. Fe atoms were deposited using an e-beam heated evaporator at a rate of approximately 0.1$\AA$/minute. A quartz micro-balance supplied by Leybold-Inficon with sub-Angstrom sensitivity was used to calibrate the iron deposition flow rate. Adsorption of NO$_2$ molecules on Ca$_x$Bi$_{2-x}$Se$_3$ was achieved by controlled \emph{in situ} exposures under static flow mode, with care to minimize photon exposure of the adsorbed surface. Large single crystals of Ca$_x$Bi$_{2-x}$Se$_3$ were grown using methods described in Ref. \cite{HorCa}. Surface and bulk state band calculations were performed for comparison with the experimental data, using the LAPW method implemented in the WIEN2K package \cite{wien2k}. Details of the calculation are identical to those described in Ref. \cite{MatthewNatPhys}.

\textbf{Corresponding author:}

Correspondence and requests for materials should be addressed to M.Z.H. (mzhasan@Princeton.edu).

\textbf{Acknowledgements:}

We acknowledge helpful discussions with R.R. Biswas and D. Haldane. We are grateful for beamline support from S. Mo.

\newpage

\begin{figure*}[t]
\includegraphics[width = 12cm]{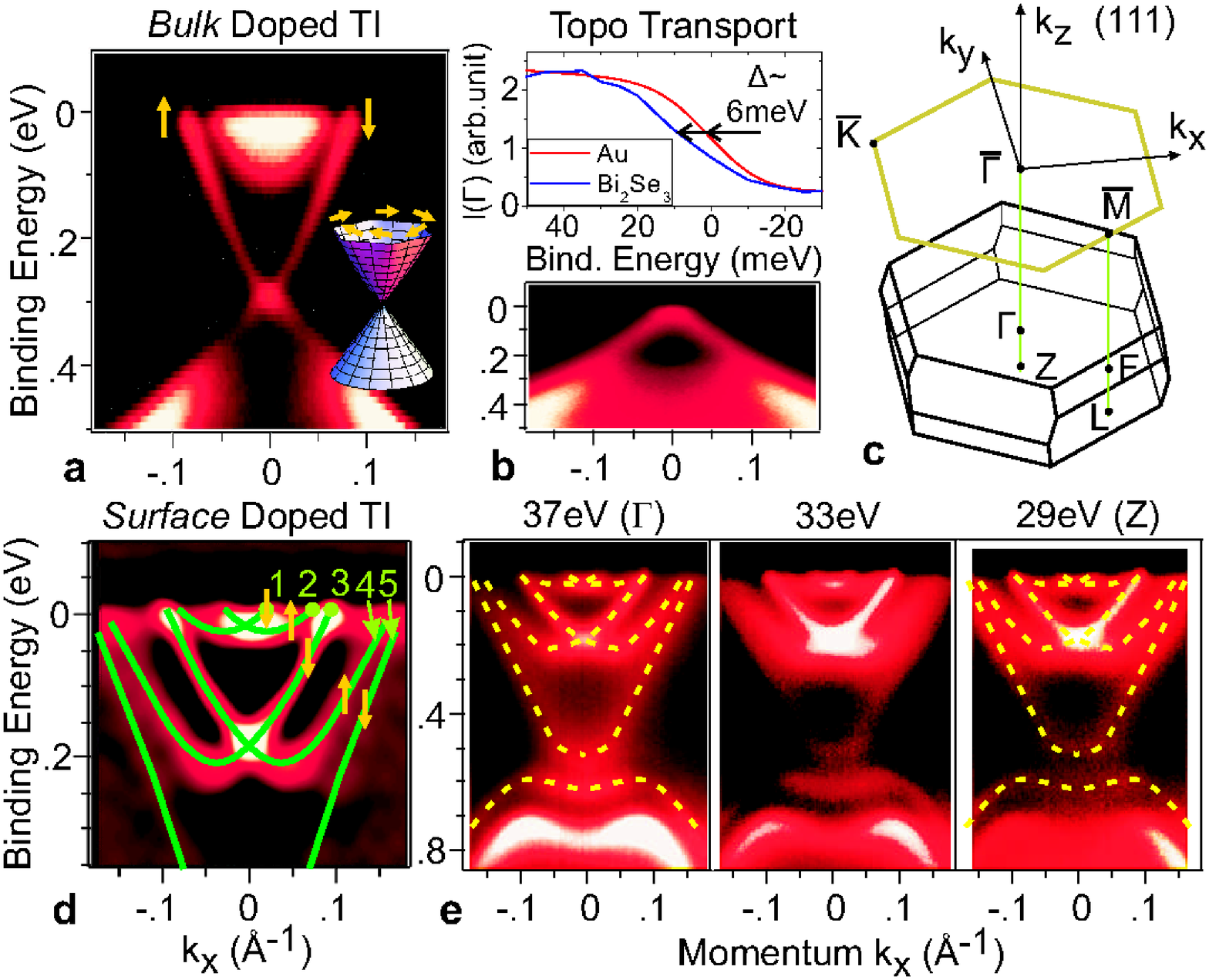}
\caption{{\bf{Iron deposition strongly modifies the topological surface}}:
\textbf{a}, Uniformly electron doped Bi$_2$Se$_3$ has a single surface state Dirac cone. \textbf{b}, When the surface chemical potential of as-grown Bi$_2$Se$_3$ is lowered to the Dirac point by NO$_2$ deposition, we observe a slight gap in the leading edge of ARPES intensity. \textbf{c}, The hexagonal surface Brillouin zone of Bi$_2$Se$_3$ is drawn above a diagram of the three dimensional bulk Brillouin zone. \textbf{d}, A second derivative image of new surface states in Bi$_2$Se$_3$ (Sample $\#$1) after surface iron deposition is labeled with numerically predicted spin texture from Fig. 3b. \textbf{e}, Low energy features from (\textbf{d}) have no z-axis momentum dispersion as seen from the data taken with varying incident photon energy (37 eV to 29 eV) confirming the two dimensional character of the state.}
\end{figure*}

\begin{figure*}[t]
\includegraphics[width = 16cm]{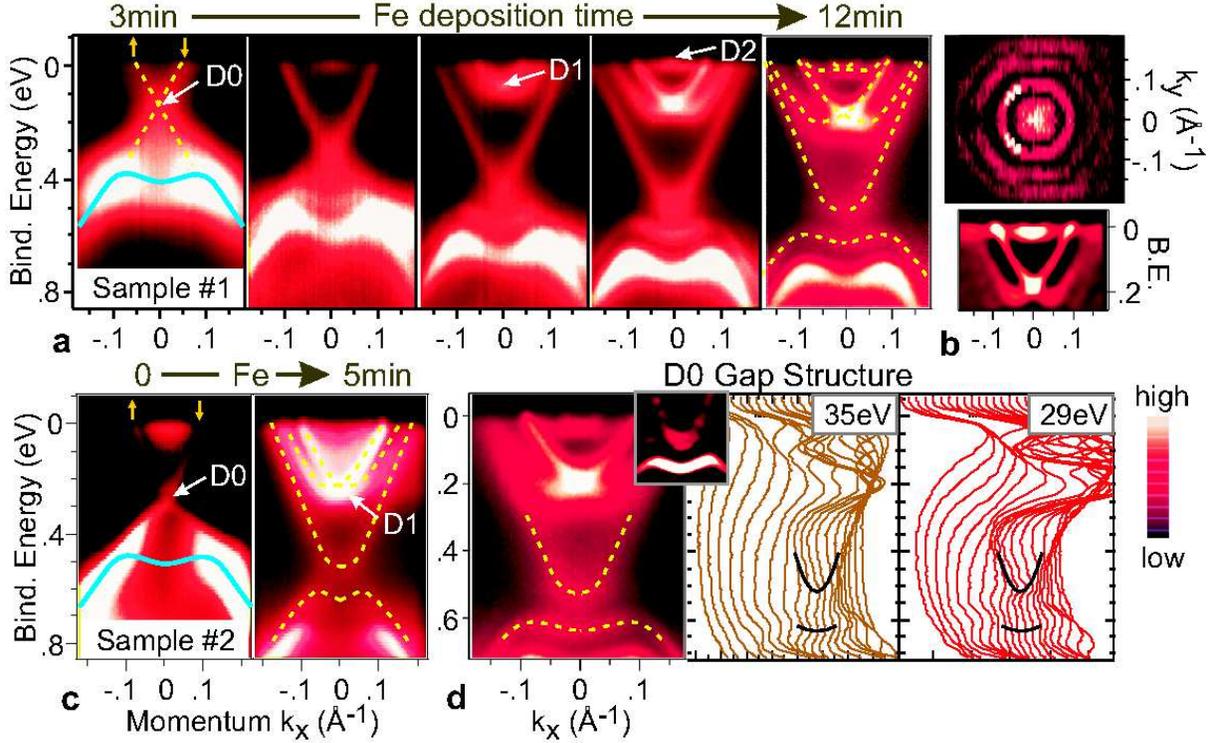}
\caption{{\bf{Iron doping creates five Dirac cones}}: \textbf{a}, Progressive surface doping of optimally insulating Ca$_x$Bi$_{2-x}$Se$_3$ (Sample $\#$1) causes the successive appearance of two new Dirac points (``D1" and ``D2"). \textbf{b}, Second derivative images of the Fermi surface and low energy surface states are shown for Fe-deposited Sample $\#$1. \textbf{c}, Iron deposition on as-grown n-type Bi$_2$Se$_3$ (Sample $\#$2) results in a similar final spectrum to panel (\textbf{a}), but with only a single new Dirac point. \textbf{d}, The iron-induced D0 gap is observed at different binding energies for Sample $\#1$ and (inset) in a second derivative image. Existence of the surface band-gap is not sample dependent.}
\end{figure*}

\begin{figure*}[t]
\includegraphics[width = 16cm]{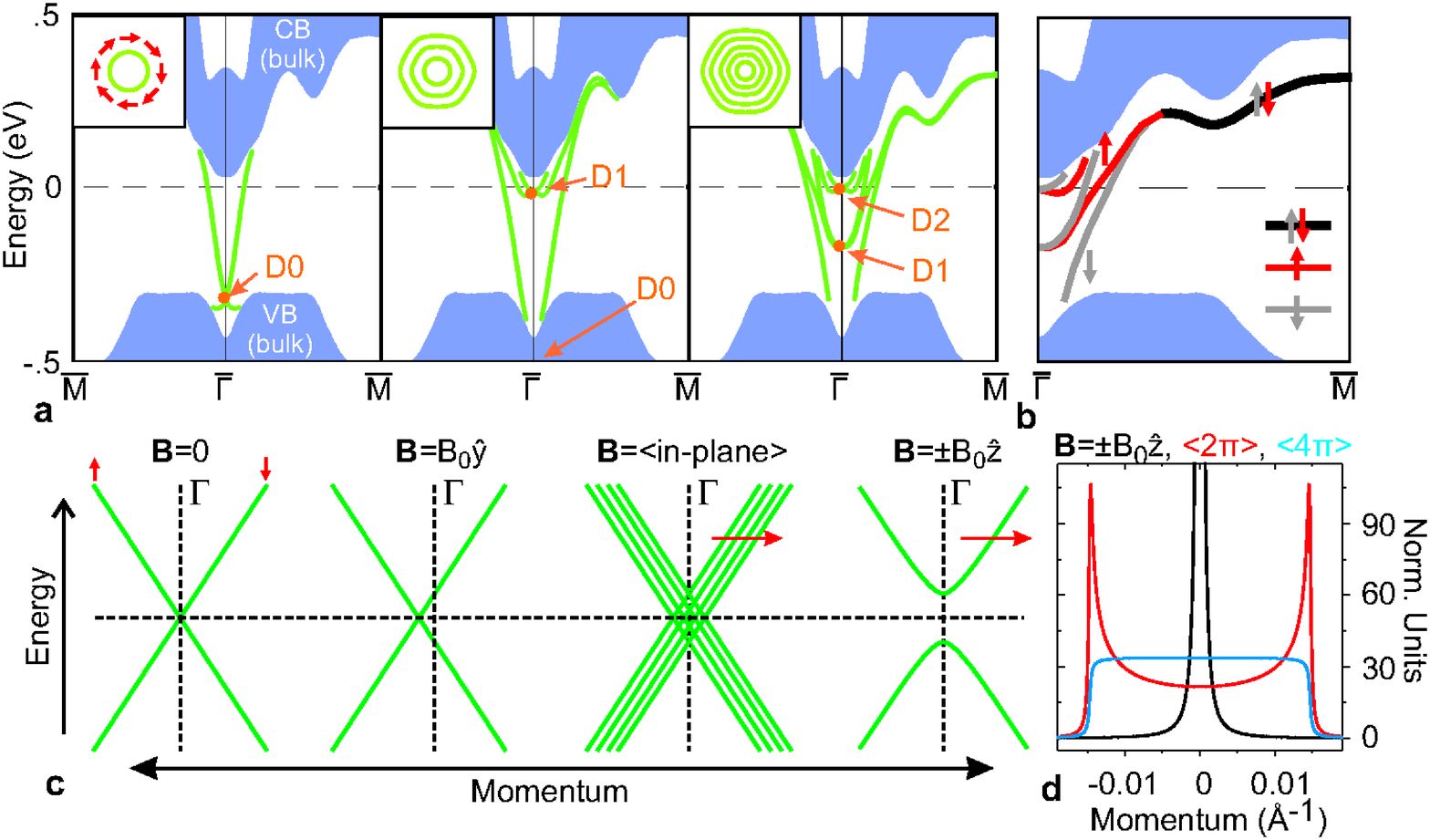}
\caption{{\bf{Surface state simulations}}: \textbf{a}, A first principles GGA model of Bi$_2$Se$_3$ surface band structure (at left) exhibits new Dirac points when orbital energies are lowered by 0.68eV on (center) the first and (right) the first two quintuple layers of the lattice. Surface state energy changes nonlinearly with this perturbation. Inset schematics show the Fermi surface evolution (not to scale). \textbf{b}, Bands from panel (\textbf{a},right) are labeled with $\uparrow$ for right handed and $\downarrow$ for left handed chiral spin texture. \textbf{c}, The effect of Zeeman coupling on a surface state Dirac cone defined by a Rashba Hamiltonian is considered for different magnetic field orientations, based on the surface state Hamiltonian in Ref. \cite{FerroSplitting}. Red arrows illustrate the path of a constant energy ARPES measurement, corresponding to profiles in panel (\textbf{d}). \textbf{d}, The domain-averaged momentum axis distribution of a single surface state band is shown for a surface populated with (black) out-of-plane ferromagnetic domains, (red) in-plane domains and (blue) domains with arbitrary 3D orientation, assuming a band velocity of 3.4eV$\cdot$A and magnetic interactions strong enough to open a 0.1eV gap when oriented along the z-axis.}
\end{figure*}

\begin{figure*}[t]
\includegraphics[width = 12cm]{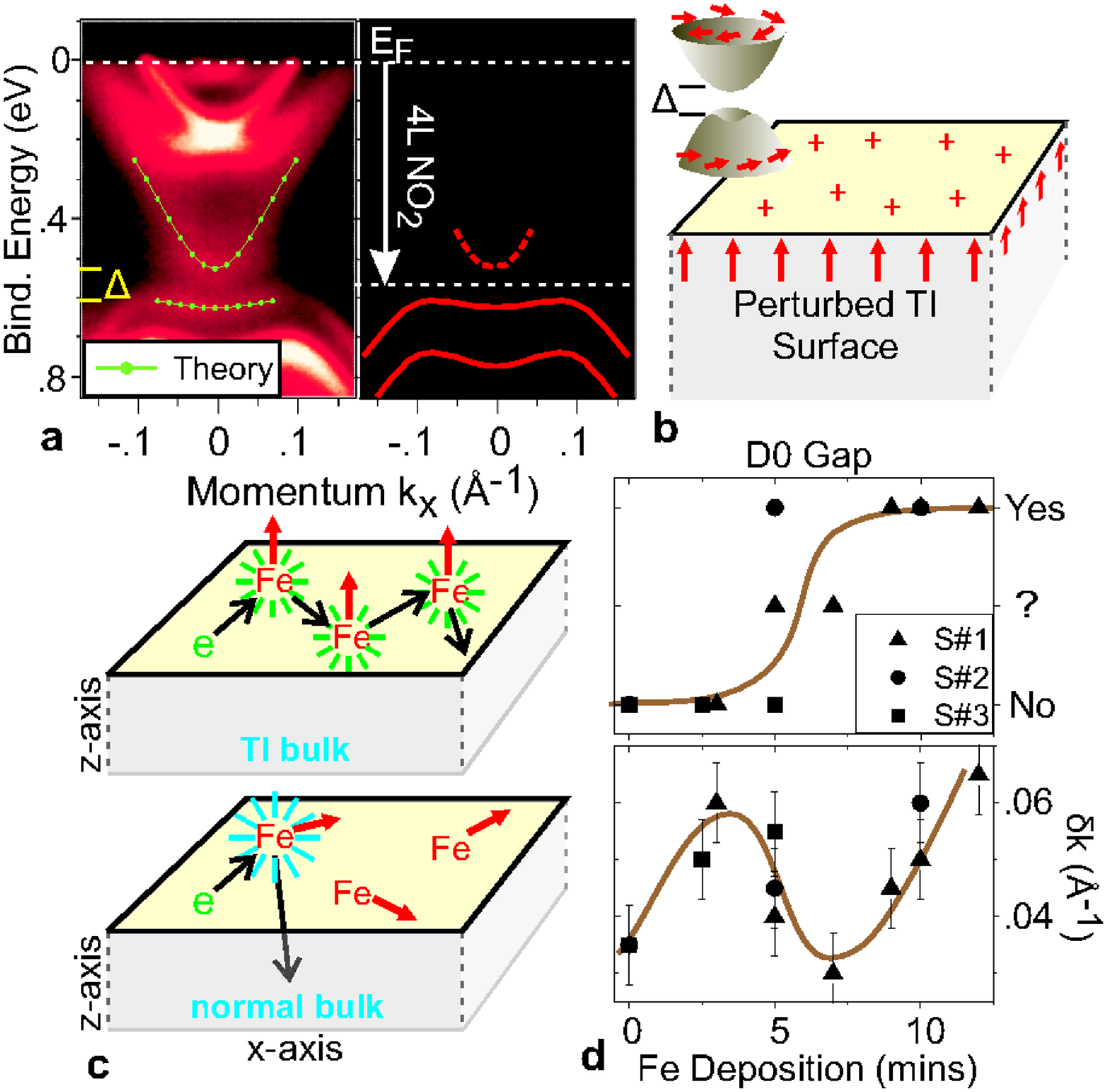}
\caption{{\bf{Surface evolution under disorder and magnetism}}: \textbf{a}, (left) A z-axis magnetic perturbation causes (green) GGA predicted surface states to conform to the iron doped dispersion. (right) A surface insulator phase can be obtained by adding $\sim$4 Langmuirs \cite{DavidTunable} of non-magnetic NO$_2$ on the surface subsequent to Fe deposition. \textbf{b}, A cartoon illustrates Fe deposition on the TI surface. \textbf{c}, (black arrows) Electron trajectories in TIs adhere to the surface, helping to mediate 2D magnetic interactions. Red arrows represent Fe spin. \textbf{d}, The development of a gap at the lowest energy (``D0") Dirac point is compared with (bottom) the half maximum peak width ($\delta$k) of connected states in the D0 upper Dirac cone. Error bars represent estimated uncertainty based on the results of different fitting techniques.}
\end{figure*}


\begin{thebibliography}[

\bibitem{hasan} Hasan, M. Z. $\&$ Kane, C. L. Topological insulators. \emph{Rev. Mod. Phys.} \textbf{82}, 3045 (2010).
\bibitem{Intro} Moore, J. E. Topological insulators: The next generation. \emph{Nature Physics} \textbf{5}, 378-380 (2009).
\bibitem{TIbasic} Fu, L., Kane, C. L., $\&$ Mele, E. J. Topological insulators in three dimensions. \emph{Phys. Rev. Lett.} \textbf{98}, 106803 (2007).
\bibitem{MooreandBal} Moore, J. E. $\&$ Balents, L. Topological invariants of time-reversal-invariant band structures. \emph{Phys. Rev. B} \textbf{75}, 121306(R) (2007).
\bibitem{DavidNat1} Hsieh, D. \emph{et al.} A topological Dirac insulator in a quantum spin Hall phase. \emph{Nature (London)} \textbf{452}, 970-974 (2008).
\bibitem{DavidScience} Hsieh, D. \emph{et al.} Observation of unconventional quantum spin textures in topological insulators. \emph{Science} \textbf{323}, 919-922 (2009).
\bibitem{DavidTunable} Hsieh, D. \emph{et al.} A tunable topological insulator in the spin helical Dirac transport regime. \emph{Nature (London)} \textbf{460}, 1101-1105 (2009).
\bibitem{MatthewNatPhys} Xia, Y. \emph{et al.} Observation of a large-gap topological-insulator class with a single Dirac cone on the surface. \emph{Nature Physics} \textbf{5}, 398-402 (2009).
\bibitem{MatthewPreprint} Fe-doping of Bi2Se3 was reported (2008) by Xia, Y. et al. Electrons on the surface of Bi2Se3 form a topologically-ordered two dimensional gas with a non-trivial Berry's phase. Preprint at http://arxiv.org/abs/0812.2078 (2008). Further details in the SI section of Hsieh et al. Nature (London) 460, 1101 (2009).
\bibitem{HorCa} Hor, Y. S. \emph{et al.}, p-type Bi$_2$Se$_3$ for topological insulator and low temperature thermoelectric applications. \emph{Phys. Rev. B} \textbf{79}, 195208 (2009).
\bibitem{HorMn} Hor, Y. S. \emph{et al.}, Development of ferromagnetism in the doped topological insulator Bi$_{2-x}$Mn$_x$Te$_3$. \emph{Phys. Rev. B} \textbf{81}, 195203 (2010)
\bibitem{YazdaniBack} Roushan, P. \emph{et al.} Topological surface states protected from backscattering by chiral spin texture. \emph{Nature} \textbf{460}, 1106-1109 (2009).

\bibitem{Biswas} Biswas, R. R. $\&$ Balatsky, A. V. Impurity-induced states on the surface of 3D topological insulators. \emph{Phys Rev B} \textbf{81}, 233405 (2010).
\bibitem{FerroSplitting} Garate, I. $\&$ Franz, M. Inverse spin-galvanic effect in a topological-insulator/ferromagnet interface. \emph{Phys. Rev. Lett.} \textbf{104}, 146802 (2010).
\bibitem{KaneDevice} Fu, L. $\&$ Kane, C. L. Probing neutral Majorana fermion edge modes with charge transport. \emph{Phys. Rev. Lett.} \textbf{102}, 216403 (2009).
\bibitem{palee} Law, K. T.,  Lee, P. A. $\&$ Ng, T. K. Majorana Fermion Induced Resonant Andreev Reflection. \emph{Phys. Rev. Lett.} \textbf{103}, 237001 (2009).
\bibitem{ZhangDyon} Qi X.-L. \emph{et al.} Inducing a magnetic monopole with topological surface states. \emph{Science} \textbf{323}, 1184-1187 (2009)
\bibitem{ExcitCapacitor} Seradjeh, B., Moore, J.E. $\&$ Franz, M. Exciton condensation and charge fractionalization in a topological insulator film. \emph{Phys. Rev. Lett.} \textbf{103}, 066402 (2009).
\bibitem{MacDonaldKerr} Tse, W.-K. $\&$ MacDonald, A.H. Giant Magneto-Optical Kerr Effect and Universal Faraday Effect in Thin-Film Topological Insulators. \emph{Phys. Rev. Lett.} \textbf{105}, 057401 (2010).
\bibitem{ZhangPred} Zhang, H. \emph{et al.} Model Hamiltonian for topological insulators \emph{Phys. Rev. B} \textbf{82}, 045122 (2010)
\bibitem{dhlee} Lee, D.-H. Surface States of Topological Insulators: The Dirac Fermion in Curved Two-Dimensional Spaces. \emph{Phys. Rev. Lett.} \textbf{103}, 196804 (2009).
\bibitem{FuHexagonal} Fu, L. Hexagonal warping effects in the surface states of topological insulator Bi$_2$Te$_3$. \emph{Phys. Rev. Lett.} \textbf{103}, 266801 (2009)
\bibitem{FuNew} Fu, L. $\&$ Berg, E. Odd-parity topological superconductors: theory and application to Cu$_x$Bi$_2$Se$_3$. Preprint at $\langle$http://arxiv.org/abs/0912.3294$\rangle$ (2009).

\bibitem{WrayCuBiSe} Wray, L. \emph{et al.} Observation of unconventional band topology in a superconducting doped topological insulator, Cu$_x$-Bi$_2$Se$_3$: Topological Superconductor or non-Abelian superconductor? Preprint at $\langle$http://arxiv.org/abs/0912.3341$\rangle$ (2009).
\bibitem{MatthewTuneBiTe} Xia, Y. \emph{et al.} Systematic control of surface Dirac fermion density on topological insulator Bi$_2$Te$_3$. Preprint at $\langle$http://arxiv.org/abs/0907.3089$\rangle$ (2009).
\bibitem{YeHelix} Ye, F. \emph{et al.} Spin helix of magnetic impurities in two-dimensional helical metal. \emph{Europhys. Lett.} \textbf{90}, 47001 (2010).
\bibitem{IronValence} Kawaminami, M. and Okazaki, A. Neutron Diffraction Study of Fe$_7$Se$_8$. II. \emph{J. Phys. Soc. Japan} \textbf{29}, 649-655 (1970).

\bibitem{zhangTImagImp} Liu, Q. \emph{et al.} Magnetic impurities on the surface of a topological insulator. \emph{Phys. Rev. Lett.} \textbf{102}, 156603 (2009).
\bibitem{adatomExchange} Wahl, P. \emph{et al.} Exchange Interaction between Single Magnetic Adatoms. \emph{Phys. Rev. Lett.} \textbf{98}, 056601 (2007).
\bibitem{MerminWagner} Mermin, N. D. $\&$ Wagner, H. Absence of ferromagnetism or antiferromagnetism in one- or two-dimensional isotropic Heisenberg models. \emph{Phys. Rev. Lett.} \textbf{17}, 1133-1136 (1966).
\bibitem{UltrathinMag} Zhang, R.-J. $\&$ Willis, R. F. Thickness-dependent Curie temperatures of ultrathin magnetic films: effect of the range of spin-spin interactions. \emph{Phys. Rev. Lett.} \textbf{86}, 2665-2668 (2001).

\bibitem{wien2k} Blaha, P. \emph{et al. Computer Code WIEN2K} (Vienna University of Technology, Vienna, 2001).

\end{thebibliography}
\end{document}